\documentclass{article}
\usepackage{amsmath, amssymb, amsfonts}
\title{On a modified Rindler geometry} 
\author{Hristu Culetu, \\Ovidius University, Dept.of Physics and Electronics, \\ Mamaia Avenue 124, 900527 Constanta, Romania, \footnote {e-mail : hculetu@yahoo.com}}

\begin{document}
\numberwithin{equation}{section}
\pagenumbering{arabic}
\maketitle
\newcommand{\fv}{\boldsymbol{f}}
\newcommand{\tv}{\boldsymbol{t}}
\newcommand{\gv}{\boldsymbol{g}}
\newcommand{\OV}{\boldsymbol{O}}
\newcommand{\wv}{\boldsymbol{w}}
\newcommand{\WV}{\boldsymbol{W}}
\newcommand{\NV}{\boldsymbol{N}}
\newcommand{\hv}{\boldsymbol{h}}
\newcommand{\yv}{\boldsymbol{y}}
\newcommand{\RE}{\textrm{Re}}
\newcommand{\IM}{\textrm{Im}}
\newcommand{\rot}{\textrm{rot}}
\newcommand{\dv}{\boldsymbol{d}}
\newcommand{\grad}{\textrm{grad}}
\newcommand{\Tr}{\textrm{Tr}}
\newcommand{\ua}{\uparrow}
\newcommand{\da}{\downarrow}
\newcommand{\ct}{\textrm{const}}
\newcommand{\xv}{\boldsymbol{x}}
\newcommand{\mv}{\boldsymbol{m}}
\newcommand{\rv}{\boldsymbol{r}}
\newcommand{\kv}{\boldsymbol{k}}
\newcommand{\VE}{\boldsymbol{V}}
\newcommand{\sv}{\boldsymbol{s}}
\newcommand{\RV}{\boldsymbol{R}}
\newcommand{\pv}{\boldsymbol{p}}
\newcommand{\PV}{\boldsymbol{P}}
\newcommand{\EV}{\boldsymbol{E}}
\newcommand{\DV}{\boldsymbol{D}}
\newcommand{\BV}{\boldsymbol{B}}
\newcommand{\HV}{\boldsymbol{H}}
\newcommand{\MV}{\boldsymbol{M}}
\newcommand{\be}{\begin{equation}}
\newcommand{\ee}{\end{equation}}
\newcommand{\ba}{\begin{eqnarray}}
\newcommand{\ea}{\end{eqnarray}}
\newcommand{\bq}{\begin{eqnarray*}}
\newcommand{\eq}{\end{eqnarray*}}
\newcommand{\pa}{\partial}
\newcommand{\f}{\frac}
\newcommand{\FV}{\boldsymbol{F}}
\newcommand{\ve}{\boldsymbol{v}}
\newcommand{\AV}{\boldsymbol{A}}
\newcommand{\jv}{\boldsymbol{j}}
\newcommand{\LV}{\boldsymbol{L}}
\newcommand{\SV}{\boldsymbol{S}}
\newcommand{\av}{\boldsymbol{a}}
\newcommand{\qv}{\boldsymbol{q}}
\newcommand{\QV}{\boldsymbol{Q}}
\newcommand{\ev}{\boldsymbol{e}}
\newcommand{\uv}{\boldsymbol{u}}
\newcommand{\KV}{\boldsymbol{K}}
\newcommand{\ro}{\boldsymbol{\rho}}
\newcommand{\si}{\boldsymbol{\sigma}}
\newcommand{\thv}{\boldsymbol{\theta}}
\newcommand{\bv}{\boldsymbol{b}}
\newcommand{\JV}{\boldsymbol{J}}
\newcommand{\nv}{\boldsymbol{n}}
\newcommand{\lv}{\boldsymbol{l}}
\newcommand{\om}{\boldsymbol{\omega}}
\newcommand{\Om}{\boldsymbol{\Omega}}
\newcommand{\Piv}{\boldsymbol{\Pi}}
\newcommand{\UV}{\boldsymbol{U}}
\newcommand{\iv}{\boldsymbol{i}}
\newcommand{\nuv}{\boldsymbol{\nu}}
\newcommand{\muv}{\boldsymbol{\mu}}
\newcommand{\lm}{\boldsymbol{\lambda}}
\newcommand{\Lm}{\boldsymbol{\Lambda}}
\newcommand{\opsi}{\overline{\psi}}
\renewcommand{\tan}{\textrm{tg}}
\renewcommand{\cot}{\textrm{ctg}}
\renewcommand{\sinh}{\textrm{sh}}
\renewcommand{\cosh}{\textrm{ch}}
\renewcommand{\tanh}{\textrm{th}}
\renewcommand{\coth}{\textrm{cth}}

\begin{abstract}
Following a previous idea, a curved geometry is proposed as being valid in accelerated systems, in Minkowski space. The curvature turns out to be generated by the source of the accelerated motion. An exponential factor depending on $\rho$ (the coordinate along the acceleration) and a constant length is introduced in the metric. The source stress tensor appears to represent an imperfect fluid with zero energy density but nonzero tangential pressures which do not depend on Newton's constant even for $\rho>>l_{p}$, where $l_{p}$ is the Planck length. The Komar mass is proportional to the constant acceleration $g$ and it does not depend on the choice of the value of the constant $k$ from the exponential factor. Null and timelike geodesics along the $\rho$-direction are investigated. A slight change in the metric leads to nonzero energy density and pressure along the acceleration direction, with all the energy conditions being satisfied far from the Planck world.
 \end{abstract}
 
 \section{Introduction}
It seems reasonable that the flat spacetime should have zero energy. But, as Padmanabhan \cite{TP} has pointed out, the ''reasonableness'' of any procedures in physics needs to be verified by experiment. It is a well-known result that an accelerating observer detects particles as if it were in a thermal heat bath at a temperature proportional with the proper acceleration \cite{PD, WU}. The thermal nature of the vacuum state may be viewed when it is expressed in terms of the Rindler wave functions.

Why an Unruh detector clicks if it is accelerating even in Minkowski space? On the grounds of the fact that the v.e.v. $<T_{ab}>$ is a covariant object, the answer would be ''No''. If regularized $<T_{ab}>$ vanishes in one frame (say, Minkowski frame), then it ought to vanish in all frames and, hence, in the Rindler frame. On the other hand, the detector will not have to click as long as it will never see any curvature of the spacetime. It is exactly that curvature we are dealing with in this paper.

 The proposal that an accelerated observer finds himself in a curved (AdS) spacetime with a negative cosmological constant generated by acceleration has been put forward in \cite{HC1}. In other words, inertial forces in accelerated systems arise from the curvature induced by a negative cosmological constant. In \cite{HC2} it was assumed that a Minkowski observer sees the spacetime as if it were full of Hawking wormholes \cite{SH, SW, HC3}, the line-element becoming static to a hyperbolic observer, with sharp variations of the Ricci tensor close to the horizons.

 In the present paper we follow a different route. Preserving the previous conjecture that a constant accelerating observer in Minkowski space finds actually in a curved background generated by the source of acceleration, we adjust the Rindler metric for to become curved, due to an exponential factor that maintain also the geometry regular. In Sec.2 we introduce the curved Rindler-like metric and the kinematical quantities and curvature invariants associated to the spacetime. The Sec.3 is devoted to the properties of the stress tensor as the source of curvature. The Komar mass, valid for our static geometry, is calculated in Sec.4. Sec.5 deals with the study of the null and timelike geodesics. Sec.6 treats on a small alteration of the basic metric, leading to more credible properties of the source stress tensor. We conclude in Sec.7 with some general remarks.

Throughout the paper geometrical units $G = c = \hbar = 1$ are used, unless otherwise specified.

  \section{Curved Rindler-like metric}
	Let us start with the following form of the flat Rindler line-element \cite{HC4}
		\begin{equation}
	   ds^{2} = - (1 + 2gx) dt^{2} + \frac{dx^{2}}{1 + 2gx} + dy^{2} + dz^{2},     
 \label{2.1}
 \end{equation} 
where $g>0$ is the constant acceleration along the $x$-direction and $x \geq -1/2g$ for to keep the signature $(-, +, +, +)$. The Rindler horizon is located at $x = -1/2g$. Changing the variable $x$ to $1+2gx = 2g\rho$, one obtains
 \begin{equation}
	   ds^{2} = - 2g\rho ~dt^{2} + \frac{d\rho^{2}}{2g\rho} + dy^{2} + dz^{2},     
 \label{2.2}
 \end{equation} 
with the horizon at $\rho = 0$. The metric (2.2) is, of course, flat. But our goal is to look for a curved metric, the curvature being generated, in our view, by the agent who set the object in accelerating motion. Therefore, we modify the above geometry to
 \begin{equation}
	   ds^{2} = - 2g\rho e^{-\frac{k}{\rho}}~dt^{2} + \frac{e^{\frac{k}{\rho}}}{2g\rho}~d\rho^{2} + dy^{2} + dz^{2},     
 \label{2.3}
 \end{equation} 
where $k$ is a positive constant, with units of length, and $y, z$ are the transversal coordinates. It is worth noting that the $g_{tt}$ metric coefficient vanishes when $\rho \rightarrow 0$ such that $\rho = 0$ is again a horizon. The function $f(\rho) = - g_{tt} = 2g\rho e^{-k/\rho}$ is an increasing function, with a null derivative at the origin $\rho = 0$. For $k = 0$ or $\rho>>k$ we get the flat Rindler metric. 
  We now consider a static observer in the geometry (2.3) having a velocity vector field
	  \begin{equation}
  u^{b} = \left(\frac{1}{\sqrt{2g\rho}}~e^{\frac{k}{2\rho}}, 0, 0, 0 \right),~~~ u^{b}u_{b} = -1,  
 \label{2.4}
 \end{equation} 
whence one finds the only nonzero component of the covariant acceleration $a^{b}= u^{a}\nabla_{a}u^{b}$ 
	  \begin{equation}
  a^{\rho} = g \left(1 + \frac{k}{\rho}\right) e^{-\frac{k}{\rho}}.
 \label{2.5}
 \end{equation} 
$a^{\rho}$ is an increasing function of $\rho$, from zero when $\rho \rightarrow 0$ to $g$ when $\rho \rightarrow \infty$, having an inflexion point at $\rho = k/3$, with $a^{\rho}(k/3) = 4g/e^{3}$. Its null value at the horizon seems unrealistic, excepting the case when the constant $k$ is supposed to have a microscopic value.

The proper acceleration reads
  \begin{equation}
A \equiv \sqrt{a^{b}a_{b}} = \sqrt{\frac{g}{2\rho}}\left(1 + \frac{k}{\rho}\right) e^{-\frac{k}{2\rho}}.
 \label{2.6}
 \end{equation}
When $\rho >>k$, we get $a^{\rho} = g$ and $A = \sqrt{g/2\rho}$, namely the well-known values from the flat Rindler metric. However, when $\rho \rightarrow 0$ (the horizon), the both $a^{\rho}$ and $A$ approach zero, very different w.r.t. the standard Rindler situation. The reason is the exponential factor, which gives also a vanishing surface gravity 
  \begin{equation}
\kappa = (\sqrt{a^{b}a_{b}} ~\sqrt{-g_{tt}})|_{\rho = 0} = a^{\rho}|_{\rho = 0} = 0.
 \label{2.7}
 \end{equation}
As long as the curvature invariants are concerned, we get
  \begin{equation}
	R^{a}_{~a} = -\frac{2gk^{2}}{\rho^{3}}~e^{-\frac{k}{\rho}},~~~K = \frac{4g^{2}k^{6}}{\rho^{6}}~e^{-\frac{2k}{\rho}},
 \label{2.8}
 \end{equation}
where $R^{a}_{~a}$ is the scalar curvature and $K = R^{abcd}R_{abcd}$ is the Kretschmann invariant ($a, b, c, d$ run from $t$ to $z$). One notices that both invariants are regular at the horizon $\rho = 0$ and at infinity (or $\rho >>k$) where they are vanishing (the metric acquires the standard Rindler form).

Let us now make a choice of the value of the constant length $k$. A first proposal might be $k = 1/g$, which is directly related to the constant accelerating motion, with acceleration $g$. However, with that value of $k$, we need to consider $\rho >> 1/g$ for to obtain $a^{\rho} \approx g$, which is unrealistic because $1/g$ is a huge value, with a value of $g$ not very high. Therefore, our choice will be $k = l_{p}$, where $l_{p} = 10^{-33}$cm is the Planck length (the second general length at our disposal). We shall see that that value of $k$ leads to very plausible results. In other words, our change of the standard Rindler metric matters only at very short distances, close to the Planck world. With that choice of $k$ the function $a^{\rho}$ has the inflexion point already at $\rho = l_{p}/3$, where $a^{\rho}$ acquires the value $4g/e^{3} \approx g/5$.  

\section{Imperfect fluid stress tensor}
We look now for the source of curvature in the geometry (2.3), namely we need the components of the stress tensor to be inserted on the r.h.s. of Einstein's equations $G_{ab} = 8\pi T_{ab}$ for to get (2.3) as an exact solution. The only nonzero components of the Einstein tensor are given by
  \begin{equation}
	G^{y}_{y} = G^{z}_{z} = \frac{gk^{2}}{\rho^{3}}~e^{-\frac{k}{\rho}}.
 \label{3.1}
 \end{equation}
From (3.1) and Einstein's equations it is clear that the fluid is imperfect because its energy density and pressure along the $\rho$-direction are vanishing. With the velocity vector field from (2.4), $T^{a}_{~b}$ may be put in the form
  \begin{equation}
	T^{a}_{~b} = (u^{a}u_{b} - n^{a}n_{b} + \delta^{a}_{~b}) p_{t},
 \label{3.2}
 \end{equation}
where $p_{t} \equiv p_{y}, p_{z}$ are the transversal pressures and $n^{a}$ is a vector normal to the surfaces of constant $\rho$, with 
	  \begin{equation}
  n^{b} = \left(0, \sqrt{2g\rho}~e^{-\frac{k}{2\rho}}, 0, 0 \right),~~~ n^{b}n_{b} = 1.  
 \label{3.3}
 \end{equation} 
One could check, using (3.2) and (3.3), that the energy density $\epsilon = T^{a}_{~b}u^{b}u_{a} = 0$, the pressure $p_{\rho} = T^{a}_{~b}n^{b}n_{a} = 0$ and 
	\begin{equation}
	p_{t} = T^{y}_{~y} = T^{z}_{~z} = \frac{gk^{2}}{8\pi \rho^{3}}~e^{-\frac{k}{\rho}}
 \label{3.4}
 \end{equation} 
Concerning the energy conditions, one finds that all energy conditions are fulfilled, aside from the dominant energy condition (DEC), because $\epsilon <p_{t}$.

 Let us estimate the value of $p_{t}$ when we introduce $k = l_{p}$ and all fundamental constants. Eq.(3.4) yields
	\begin{equation}
	p_{t} = \frac{g\hbar}{8\pi c\rho^{3}}~e^{-\frac{l_{p}}{\rho}}.
 \label{3.5}
 \end{equation} 
It is worth noting that far from the Planck world ($\rho >>l_{p}$) one obtains $p_{t} \approx \hbar g/8\pi c\rho^{3}$. It can be very large with a high $g$ and small $\rho$, for instance of the order of nuclear or atomic distances. Moreover, with the above approximation, $p_{t}$ does not depend on $G$ but only on $\hbar$ and $c$, having so a quantum origin. For example, with $\rho = 10^{-15}cm$ and $g = 10^{20}cm/s^{2}$, we have $R^{a}_{~a} \approx -10^{-22} cm^{-2}$ and $p_{t} \approx 10^{26} erg/cm^{3}$, that is a huge pressure. On the contrary, $\rho = 1 cm$ and $g = 10^{3}cm/s^{2}$ gives us $p_{t} \approx 10^{-35} erg/cm^{3}$, a completely negligible value. We see that $p_{t}$ matters only in microphysics.\\
 Concerning the function $p_{t}(\rho)$, it is vanishing when $\rho \rightarrow 0$ or $\rho \rightarrow \infty$ and has a maximum at $\rho = l_{p}/3$, with $p_{t,max} = 27g/8\pi e^{3} l_{p} \approx g\cdot 10^{60} ergs/cm^{3}$, that is very large even with a small $g$. Nevertheless, this $p_{t,max}$ is reached for $\rho <l_{p}$. Even with very large $g$ we are unable to achieve such a small value of $\rho$. Anyway, to observe large values of $p_{t}$ or $R^{a}_{~a}$ in the accelerating system we have to perform measurements at short distances or in very short time intervals.

 \section{The Komar mass}
It is instructive to compute the Komar mass-energy $W_{K}$ \cite{TP1}
 	\begin{equation}
	W_{K} = 2 \int_{V}(T_{ab} - \frac{1}{2} g_{ab}T^{c}_{~c})u^{a} u^{b} N\sqrt{h} d^{3}x ,
 \label{4.1}
 \end{equation} 
where $V$ is a 3-volume in our static metric, $N^{2} = -g_{tt}$ is the lapse function, $h = e^{\frac{k}{\rho}}/2g\rho$ is the determinant of the spatial 3 - metric, $h_{ab} = g_{ab} + u_{a} u_{b}$, and $d^{3}x = d\rho~dy~dz$. With $u^{a}$ from (2.4) and $T_{ab}$ from (3.1), Eq.(4.1) yields
 	\begin{equation}
	W_{K}(\rho) = \frac{gk^{2}}{4\pi} \Delta y \Delta z \int^{\rho_{2}}_{\rho_{1}} \frac{1}{r^{3}} e^{-\frac{k}{r}} dr
 \label{4.2}
 \end{equation} 
where we have taken $\rho \in [\rho_{1}, \rho_{2}],~ y\in [y_{1}, y_{2}],~z\in [z_{1}, z_{2}]$, with $\Delta y = y_{2} - y_{1},~\Delta z = z_{2} - z_{1}$. An evaluation of the above integral gives us
  	\begin{equation}
	W_{K} = \frac{g}{4\pi} \Delta y \Delta z \left[\left(1 + \frac{k}{\rho_{2}}\right) e^{-\frac{k}{\rho_{2}}} - \left(1 + \frac{k}{\rho_{1}}\right) e^{-\frac{k}{\rho_{1}}}\right] 
 \label{4.3}
 \end{equation} 
  It is worth noting that $W_{K}$ represents the energy within a prism with the area of the base equal to $\Delta y \Delta z$ and height $\Delta \rho = \rho_{2} - \rho_{1}$. If we let $\rho_{1} \rightarrow 0$ and $\rho_{2} \rightarrow \infty$, the Komar mass (4.3) becomes $W_{K} = (g/4\pi)\Delta y \Delta z$, with no dependence on the constant $k = l_{p}$. When the fundamental constants are introduced in its expression, we obtain $W_{K} = (gc^{2}/4\pi G) \Delta y \Delta z$. 
	
	Let us estimate what value acquires the Komar energy for macroscopic values of the variable $\rho$. Take, for instance, $\rho_{1} = 1cm,~\rho_{2} = 10^{5}cm = 1Km$ and $g = 10^{3} cm/s^{2}$. Keeping in mind that $\rho_{1,2} >>k,~\rho_{1} << \rho_{2}$, one obtains approximately
	\begin{equation}
	W_{K} = \frac{g}{4\pi} \Delta y \Delta z~ k^{2}\left(\frac{1}{\rho_{1}^{2}} - \frac{1}{\rho_{2}^{2}}\right) \approx \frac{g}{4\pi} \Delta y \Delta z \frac{k^{2}}{\rho_{1}^{2}} = 10^{-36}ergs,
 \label{4.4}
 \end{equation} 
a completely negligible value. In contrast, $W_{K}$ is much larger with microscopic $\rho$, say $\rho_{1} = 10^{-15}cm, \rho_{2} = 10^{-10}cm$ and $g = 10^{20}cm/s^{2}$. Using the same approximations, (4.3) yields
	\begin{equation}
	W_{K} = \frac{g}{4\pi} \Delta y \Delta z \frac{k^{2}}{\rho_{1}^{2}} = 10^{11}ergs. 
 \label{4.5}
 \end{equation} 
 With the above values of $\rho$ we are still far from the Planck world. It is clear that, going closer to the horizon $\rho = 0$ (or closer to the Planckian distances from the horizon), $W_{K}$ grows bigger and bigger.

\section{Null and timelike geodesics}
The next task is to find the geodesics in the geometry (2.3), taking $y, z = const.$. Let us firstly start with the null geodesics along the $\rho$-direction. \\
The Lagrangian associated to the $\rho$- motion is \cite{LCMC}
 \begin{equation}
 L = 2g\rho e^{-\frac{k}{\rho}}\dot{t}^{2} - \frac{e^{\frac{k}{\rho}}}{2g\rho}~\dot{\rho}^{2} = 0 ,
 \label{5.1}
 \end{equation} 
where a dot means derivative w.r.t. the affine parameter $\lambda$. 
 The Euler-Lagrange equations read
  \begin{equation}
	\frac{\partial L}{\partial t} - \frac{d}{d\lambda}\frac{\partial L}{\partial \dot{t}} = 0,~~~~\frac{\partial L}{\partial \rho} - \frac{d}{d\lambda}\frac{\partial L}{\partial \dot{\rho}} = 0
 \label{5.2}
 \end{equation}
From (5.1) and (5.2) we find that 
  \begin{equation}
	\frac{\partial L}{\partial \dot{t}} = 4g\rho e^{-\frac{k}{\rho}} \dot{t} = E
 \label{5.3}
 \end{equation}
where $E$ is a positive constant related to the energy of the test particle. Once the expression of the $\dot{t}$ from (5.3) is introduced in (5.1), we get $\dot{\rho} = \pm E/2$, or $\rho (\lambda) = (\pm{E/2})\lambda$. From the previous relations one obtains the equation of motion
  \begin{equation}
	\frac{d\rho}{dt} = \pm {2g\rho}~ e^{-\frac{k}{\rho}} 
 \label{5.4}
 \end{equation}
 However, the equation of motion (5.4) cannot be solved exactly analitically. A series expansion leads to 
  \begin{equation}
	ln~\left(\frac{k}{\rho}\right) + \frac{k}{\rho} + \frac{1}{2\cdot 2!} \frac{k^{2}}{\rho^{2}} + \frac{1}{3\cdot 3!} \frac{k^{3}}{\rho^{3}} +...= \pm {2gt} + \beta,
 \label{5.5}
 \end{equation}
where $\beta$ is a constant of integration.

One could check that, with the above expressions of $\dot{t}$ and $\dot{\rho}$, the 2nd equation from (5.2) is obeyed. In addition, the null vector 
  \begin{equation}
	v^{a} = (\dot{t}, \dot{\rho}, 0, 0) = \left(\frac{E}{4g\rho}~e^{\frac{k}{\rho}}, \frac{E}{2}, 0, 0\right)
 \label{5.6}
 \end{equation}
is tangent to the null geodesics, i.e. $a^{b}= v^{a}\nabla_{a}v^{b} = 0$. 

Let us look now for the timelike geodesics along the $\rho$-direction. The expression for $\dot{t} = dt/d\tau$, where $\tau$ is the proper time, is the same as in (5.3) but we have now $L = 1$, so that for $\dot{\rho} = d\rho/d\tau$ we get
 \begin{equation}
\dot{\rho}^{2} = \frac{E^{2}}{4} - 2g\rho~e^{-\frac{k}{\rho}},
 \label{5.7}
 \end{equation} 
whence one obtains
  \begin{equation}
\frac{d\rho}{dt} = \pm {\frac{2g\rho}{E}} e^{-\frac{k}{\rho}} \sqrt{E^{2} - 8g\rho e^{-\frac{k}{\rho}}}.     
 \label{5.8}
 \end{equation} 
From (5.8) it is clear that we have to consider $\rho e^{-k/\rho}\leq E^{2}/8g$. In other words, $\rho$ should be smaller than some $\rho_{0}$ to which the expression under the square root from (5.8) vanishes.

 \section{Source with nonzero energy density}
The metric (2.3) is generated by the stress tensor (3.2) with $p_{t}$ given by (3.4) and with zero energy density and pressure in the $\rho$-direction, a property which seems inappropriate. Therefore, we propose in this section a geometry slightly different from (2.3), considering that there exists a dependence of the transversal geometry on the $\rho$-coordinate. We so have
 \begin{equation}
	   ds^{2} = - 2g\rho e^{-\frac{k}{\rho}}~dt^{2} + \frac{e^{\frac{k}{\rho}}}{2g\rho}~d\rho^{2} + e^{-\frac{k}{\rho}} (dy^{2} + dz^{2}),     
 \label{6.1}
 \end{equation} 
Although the acceleration (2.5) and surface gravity (2.7) do not change, the new curvature invariants are given by
 \begin{equation}
	R^{a}_{~a} = -\frac{4gk}{\rho^{2}}\left(1 - \frac{9k}{4\rho}\right)e^{-\frac{k}{\rho}},~~~K = \frac{20~g^{2}k^{2}}{\rho^{4}}\left(1 - \frac{k}{\rho} + \frac{3k^{2}}{\rho^{2}}\right)e^{-\frac{2k}{\rho}},
 \label{6.2}
 \end{equation} 
with $K>0$ for any $\rho$.

Using Maple package one finds the The Einstein tensor from the Einstein equations and then the components of the stress tensor, which are given by
  \begin{equation}
	\begin{split}
	8\pi T^{t}_{~t} = -\frac{3gk}{\rho^{2}} \left(1 - \frac{5k}{6\rho}\right) e^{-\frac{k}{\rho}},~~~8\pi T^{\rho}_{~\rho} = \frac{gk}{\rho^{2}} \left(1 + \frac{3k}{2\rho}\right) e^{-\frac{k}{\rho}}\\ 8\pi T^{y}_{~y} =  8\pi T^{z}_{~z} = -\frac{gk}{\rho^{2}} \left(1 - \frac{5k}{2\rho}\right) e^{-\frac{k}{\rho}}
 \label{6.3}
\end{split}
 \end{equation} 
Instead of the form (3.2) of the energy-momentum tensor, we now assume that 
  	  \begin{equation}
	 T_{ab} = (\bar{p_{t}} + \bar{\epsilon}) u_{a} u_{b} + \bar{p_{t}} g_{ab} + (\bar{p_{\rho}} - \bar{p_{t}}) n_{a}n_{b},
 \label{6.4}
 \end{equation}
where the bar sign signifies the new values of the energy density and pressures, which are given by
  \begin{equation}
	\begin{split}
	8\pi \bar{\epsilon} = \frac{3gk}{\rho^{2}} \left(1 - \frac{5k}{6\rho}\right) e^{-\frac{k}{\rho}},~~~~~8\pi \bar{p_{\rho}} = \frac{gk}{\rho^{2}} \left(1 + \frac{3k}{2\rho}\right) e^{-\frac{k}{\rho}},\\ 8\pi \bar{p_{t}} = -\frac{gk}{\rho^{2}} \left(1 - \frac{5k}{2\rho}\right) e^{-\frac{k}{\rho}}
 \label{6.5}
\end{split}
 \end{equation} 
As long as the energy conditions are concerned, one observes that the WEC, SEC and NEC are obeyed if $\rho >2k$ but DEC is obeyed when $\rho >5k/2$. In other words, all energy conditions are satisfied if $\rho >5k/2 = 5l_{p}/2$. Taking $\rho >>l_{p}$ assures that all energy conditions are respected. That is of course valid even in atomic or nuclear distances, namely far from the Planck world. Compared to the situation below Eq.(3.5), for the scalar curvature we get now $R^{a}_{~a} \approx 4\cdot 10^{-4} cm^{-2}$, a much larger value; same valid for the pressures. From (6.5) it is clear that $\bar{\epsilon}$ and $\bar{p_{t}}$ change sign close to $\rho = k$.

For $\rho >>l_{p}$, with all the universal constants included, we can write
  \begin{equation}
	8\pi \bar{\epsilon} \approx \frac{c^{4}}{G} \frac{3gl_{p}}{c^{2}} \frac{1}{\rho^{2}} = \frac{3gm_{p}}{\rho^{2}},
 \label{6.6}
 \end{equation}
where $m_{p} = \sqrt{c\hbar /G}\approx 10^{-5}$ grams is the Planck mass. It is worth noting that $\bar{\epsilon}$ depends on fundamental constants only through $m_{p}$.

To obtain the expression of the new Komar mass we apply again (4.1). We have now
 	\begin{equation}
	(T_{ab} - \frac{1}{2} g_{ab}T^{c}_{~c})u^{a} u^{b} = \frac{1}{2} (\bar{\epsilon} + \bar{p_{\rho}} + 2\bar{p_{t}}) = \frac{gk}{\rho^{2}} \left(1 + \frac{2k}{\rho}\right) e^{-\frac{k}{\rho}}.
 \label{6.7}
 \end{equation} 
With $\sqrt{h} = e^{-\frac{k}{2\rho}}/\sqrt{2g\rho}$, (4.1) and (6.7) yield us
 	\begin{equation}
	\begin{split}
	W_{K}(\rho) = \frac{gk}{4\pi} \Delta y \Delta z \int^{\rho_{2}}_{\rho_{1}} \frac{1}{r^{2}}\left(1 + \frac{2k}{r}\right) e^{-\frac{2k}{r}} dr =  \\
		 \frac{g}{4\pi} \Delta y \Delta z \left[\left(1 + \frac{k}{\rho_{2}}\right) e^{-\frac{2k}{\rho_{2}}} - \left(1 + \frac{k}{\rho_{1}}\right) e^{-\frac{2k}{\rho_{1}}}\right] 
 \label{6.8}
\end{split}
 \end{equation} 
Compared to the previous expression (4.3), one notices that the only difference is the exponential $e^{-\frac{2k}{\rho}}$ instead of $e^{-\frac{k}{\rho}}$.

\section{Conclusions}
A modification of the Rindler metric by means of an exponential factor renders it curved, especially in microphysics. That factor depends on the coordinate $\rho$ along the acceleration direction and on a constant which has been chosen to be the Planck length $l_{p}$. The geometry has a horizon at $\rho = 0$, with zero surface gravity. The source stress tensor represents an imperfect fluid with zero energy density but nonzero tangential pressures which could take large values even when $\rho >>l_{p}$. 

We found instructive to calculate the Komar mass. Its value is proportional to the constant acceleration $g$ but does not depend on the Planck constant $\hbar$. As a next goal we studied the null and timelike geodesics along the $\rho$ -direction but we were not able to find the analytical form of the geodesic equations due to the exponential factor from the spacetime.

  A slight change in the metric (with the transversal geometry depending on the $\rho$-coordinate) leads to nonzero energy density and pressure along the acceleration direction, with all energy conditions being satisfied far from the Planck world.

\end{document}